\documentclass[useAMS,usenatbib]{mn2e}

\usepackage{graphicx}
\usepackage{amssymb}
\usepackage{amsmath}
\usepackage{tabularx}

\title[Cluster formation in molecular clouds]{Cluster formation in molecular clouds: I. stellar populations, star formation rates, and ionizing radiation}
\author[C.S.\ Howard, R.E.\ Pudritz, \& W.E.\ Harris ]{Corey \ S. \ Howard$^{1}$\thanks{E-mail: howardcs@mcmaster.ca}, Ralph \ E.\ Pudritz$^{1,2}$, William \ E.\ Harris$^{1}$\\
$^{1}$Department of Physics and Astronomy, McMaster University, 1280 Main St.~W, Hamilton, ON L8S 4M1, Canada\\
$^{2}$Origins Institute, McMaster University, 1280 Main St.~W, Hamilton, ON L8S 4M1, Canada}

\begin{document}
\bibliographystyle{mn2e}

\date{8 April 2013}

\pagerange{\pageref{firstpage}--\pageref{lastpage}} \pubyear{2013}

\maketitle

\label{firstpage}

\begin{abstract}
We present a model for the radiative output of star clusters in the process of star formation suitable for use in hydrodynamical simulations of radiative feedback. Gas in a clump, defined as a region whose density exceeds 10$^{4}$ cm$^{-3}$, is converted to stars via the random sampling of the Chabrier IMF. A star formation efficiency controls the rate of star formation. We have completed a suite of simulations which follow the evolution of accretion-fed clumps with initial masses ranging from 0 to 
10$^5$ M$_{\odot}$ and accretion rates ranging from 10$^{-5}$ to 10$^{-1}$ M$_{\odot}$ yr$^{-1}$. The stellar content is tracked over time which allows the aggregate luminosity, ionizing photon rate,
number of stars, and star formation rate (SFR) to be determined. For a fiducial clump of 10$^4$ M$_{\odot}$, the luminosity is $\sim$4$\times$10$^6$ L$_{\odot}$ with a SFR of roughly 3$\times$10$^{-3}$ M$_{\odot}$ yr$^{-1}$. We identify two regimes in our model. The \emph{accretion-dominated} regime obtains the majority of its gas through accretion
and is characterized by an increasing SFR while the \emph{reservoir-dominated} regime has the majority of its mass present in the initial clump with a decreasing
SFR. We show that our model can reproduce the expected number of O stars, which dominate the radiative output of the cluster. We find a nearly linear relationship between SFR and mass as seen in observations. 
We conclude that our model is an accurate and straightforward way to represent the output of clusters in hydrodynamical simulations with radiative feedback.
\end{abstract}


\section{Introduction}

\indent The clustered nature of star formation plays an important role over a wide range of spatial scales. Since a large fraction of stars, up to $\sim$90$\%$, form in a cluster environment \citep{Lada2003,Bressert2010}, understanding the cluster formation process can provide insight into the origin of the initial mass function.
On larger scales, star formation in clusters may play a central role in galactic scale feedback and the formation and destruction of giant molecular clouds (GMCs). Clusters are also interesting astrophysical objects in their own right since they can have a wide range of ages and masses
 while showing similar stellar contents.

Broadly speaking, the formation of a cluster can be thought of as the conversion into stars of a high density clump of molecular gas within a larger GMC. This process can roughly be thought of in two separate steps. Firstly, the molecular gas in a clump becomes gravitationally
unstable and fragments into protostars. Secondly, feedback mechanisms are responsible for shutting off the accretion onto the protostars and dispersing the gas in the cluster's vicinity. 

The conversion of molecular gas into fully formed stars is an inherently inefficient process \citep{Lada2003}. While feedback is certainly playing a role in the low efficiency, there are other proposed mechanisms that limit star formation. Initial turbulent velocity fields have been shown to significantly decrease the star formation efficiency per freefall time \citep{Bate2003,Bonnell2008}. However, given enough time, a turbulent molecular cloud
will be converted to stars with a 100$\%$ efficiency in the absence of other mechanisms. Magnetic fields also provide a pressure support which can significantly decrease the star formation efficiency \citep{MyersGoodman1988}. The process of feedback, however, does not just slow the star formation process but can shut off accretion to the cluster altogether. It is also important to consider because feedback affects the natal GMC since a clump and young cluster are not isolated
from their surroundings.

The interplay of turbulence, gravity, and star formation in the early stages of cluster formation has been studied in detail \citep{Klessen2001,McKee2003,Klessen2007,McKee2007}. Recent $Herschel$ studies emphasize that molecular clouds are highly filamentary \citep{Schneider2012}. Embedded young clusters in nearby molecular clouds appear at the joining points of several filaments \citep{Schneider2012,Kirk2013}. These are regions that can be fed by higher than average accretion rates \citep{Balsara, Banerjee+2006}.  
In a turbulent medium, these points are dispersed so that monolithic collapse to form a cluster does not occur. The observations of the Orion cloud suggest that subclustering occurs across large regions of the cloud \citep{Megeath2012} so that the formation of a cluster would involve the eventual merger of a significant number of subclusters.

There are several mechanisms that have been suggested as being responsible for shutting off accretion onto protostars and dispersing the remaining gas. These include stellar winds \citep{Dale2008}, radiation pressure \citep{Krumholz2012}, ionization and heating of the surrounding gas \citep{Dale2005,Peters2010,Klassen2012} 
and outflows from protostars in the presence of magnetic fields \citep{Li2006, Maury2009}. Of particular importance, and the focus of this paper, is feedback due to gas ionization and heating. Ionization is a vital process to include into numerical simulations because it produces observed HII regions. 

To fully include radiative feedback effects requires a detailed radiative transfer scheme which can be computationally intensive. Nonetheless, the effects of radiative feedback from clusters have been examined on both small and large spatial scales \citep{Dale2007,Peters2010,Bate2012,Dale2012,Klassen2012,Kim2012}. Both approaches have advantages and disadvantages. Small-scale simulations of clusters,
or multiple small clusters, simulate the formation of individual stars \citep{Bate2012}. This is advantageous because the radiative output of stars has been studied extensively through the use of stellar evolution codes. Also, since individual stars in these simulations
can be resolved, studies aimed at the origin of the IMF can be performed. However, these simulations require high resolution and are therefore computationally expensive.

Galactic scale simulations which include radiative feedback effects cannot resolve the formation of individual stars. As a compromise in these galaxy-scale models, clusters are represented as a single object with a subgrid model to represent its radiative output \citep{Tasker2011,Hopkins,Ceverino}. The clusters in these cases are typically given a fixed output that does not change with time. As an example, the luminosity of a cluster
can be determined from its mass via an averaged IMF \citep{Murray2010}. This simplified approach misses key aspects of the star formation process.

As an attempt to bridge the gap between these two types of simulations, we present a model which can be used to represent the radiative output of a star cluster. This model was produced with the ultimate goal of being integrated into hydrodynamical simulations of cluster formation
in giant molecular clouds.  

Star forming cores that form individual stars are observed to follow a mass distribution (the so-called core mass function, CMF) that follows the IMF in structure, but displaced upwards in mass by roughly a factor of 3 \citep{Alves2009,Motte2010}.  
So whatever the physical processes are in the gas that organize the CMF, we know that the outcome will resemble the IMF.  A suggested star formation efficiency of around 30\%,therefore, would give the distribution of star forming gas in a cluster forming region. 

In our model therefore, we assume that the gas that accretes onto a cluster forming region is organized in this way. As the accretion brings fresh gas into a clump, more becomes available to be distributed amongst the star forming cores. We model this by randomly sampling the available gas reservoir, drawing from an overall distribution function that is the Chabrier IMF in order to decide the masses of the most recently formed stars.
The mass spectrum of the protocluster clearly evolves with time, as gas is converted to stars from two gas sources - the initial gas mass of a clump, and the accreted mass from the external GMC. The model tracks the total cluster luminosity, number of ionizing photons, 
and the number and masses of stars contained in the cluster. These parameters can then be passed to a radiative transfer scheme to examine the effects of radiative feedback.

In section 3, we briefly highlight recent observations of embedded, star forming regions. Section 4 describes our basic model for cluster formation and is 
followed by results from a suite of simulations in section 5 which show that we are able to accurately capture the properties of young clusters. 

\section{Embedded star clusters: observations}

The earliest phases of star formation are deeply embedded in molecular gas and therefore cannot be observed at optical wavelengths. However, molecular clouds are significantly less opaque at
infrared wavelengths which allows for detailed surveys of embedded clusters with infrared telescopes such as $Herschel$. These studies indicate that embedded star formation accounts
for a large fraction of all star formation taking place in not only the Milky Way but other galaxies as well \citep{Grijs2010}. Overall, the formation of the embedded cluster phase
lasts approximately 2-4 Myr with the dissolution of the cluster through dynamical interactions and gas expulsion occuring within 10 Myr \citep{Lada2010}. 

Observational studies are also revealing the complexity of cluster forming environments. $Herschel$ observations in particular are highly filamentary \citep{Andre2011,Schneider2012}. Filaments can arise from a variety of processes including the passage of turbulent shocks in the ISM \citep{Heitsch,Balsara,Boldyrev,Pudritz2012}, gravitational collapse \citep{Hartmann,Peters2012}, and thermal instabilities \citep{Vasquez2000}. 
The largest areas of star formation occur at the junction of two or more filaments \citep{Schneider2012}. Growing clusters are fed by flow along the filaments \citep{Kirk2013} which could prolong star formation since radiation and momentum feedback can be released into the lower density regions perpendicular to the filaments. 

There is a large range of observed, embedded cluster masses and we have created our model with the goal of being able to reproduce this range. An example of a small embedded cluster is the Serpens South star forming region. This is a young cluster which currently contains
approximately 90 young stellar objects (YSOs) which are being actively fed by filamentary flows \citep{Kirk2013}. An intermediate mass example is the well-studied Orion Nebula Cluster (ONC) which is the nearest site of massive star formation. This is a young cluster (a few Myr old) with a present day mass of $\sim$ 4800 M$_{\odot}$ \citep{Hillenbrand1998} and about 2200 stars contained within a radius of 2pc from its centre \citep{Pudritz2002}.
Unlike Serpens South, the ONC is the host to massive star formation with roughly 5 stars having masses greater than 16 M$_{\odot}$ which fall in the O star range \citep{Hillenbrand1997}. As mentioned earlier, there is strong evidence for subclustering in the ONC \citep{Megeath2012}. An extremely massive example would be R136 which is the core of a ``super star cluster'' at the centre of the 30 Doradus complex. $Hubble$ $Space$ $Telescope$ imaging has revealed over 3500 stars in the centre of R136 with more than 120 of these stars being 
blue and more luminous than M$_v \sim$ -4 \citep{Massey1998}. This cluster has stellar densities which are 100-300 times more dense than any other cluster in the Milky Way or the LMC \citep{Hunter1996} and contains several stars whose masses probably exceed 100 M$_{\odot}$ \citep{Massey1998}. 

\begin{figure*}
 \begin{center}
  \includegraphics[width=0.65\linewidth]{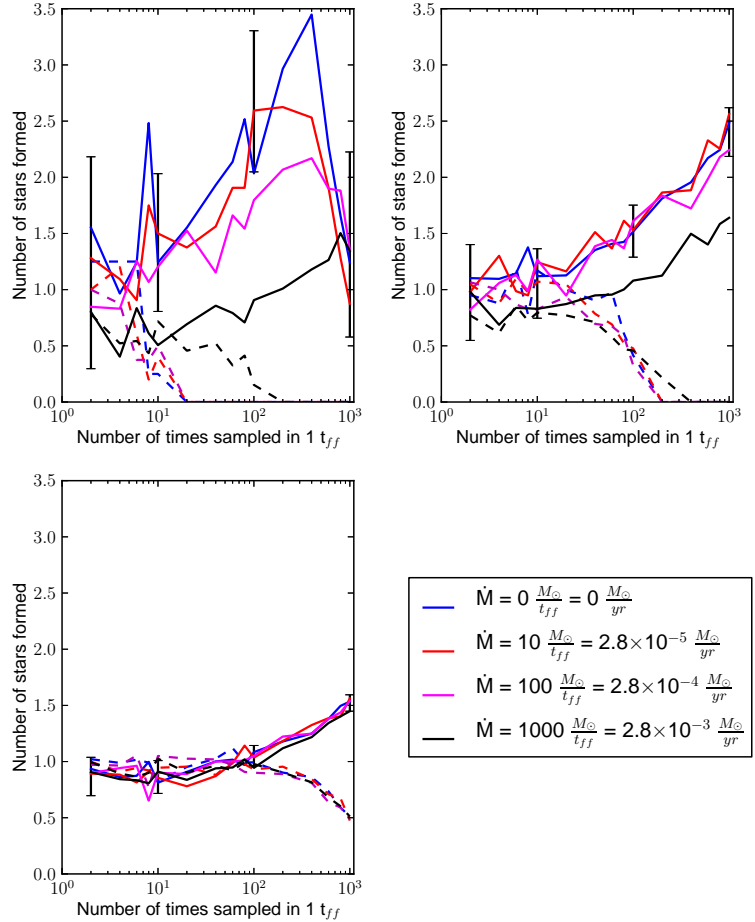}
 \end{center}
 \caption{Plots of the number of stars relative to the expected number versus sampling frequency for clusters with different initial clump masses and different constant accretion rates. The initial masses of the cluster are 100 (top left), 1000 (top right), and 10000 (bottom left) M$_{\odot}$. The solid lines represent stars less than 1 M$_{\odot}$ and dotted lines represent
 stars greater than 1 M$_{\odot}$. A subset of the points were run 100 times in order to quantify the variance between runs. The resulting average is plotted and the error bars represent 2 standard deviations.}
\end{figure*}

A high density clump of molecular gas is required to form an embedded star cluster regardless of its mass. Observations of star-forming and star-less clumps suggest that a number density of 10$^4$ cm$^{-3}$ is required to start the star forming process \citep{Lada2003}. The efficiency with which clouds are converted to stars is still in dispute. Individual
clouds in our Galaxy have a mean density of a few times 10$^2$ cm$^{-3}$ and have global star formation efficiencies ranging from 2$\%$ to 8$\%$ \citep{Kennicutt2012}. Embedded clusters have higher efficiencies between 10-30$\%$ \citep{Lada2003}. Hydrodynamical simulations indicate that the actual star formation efficiency may need to be as high at 70$\%$ to have stable, bound clusters \citep{Grijs2010}.

An important quantity which can be compared to our work is the star formation rate (SFR). The star formation rate in local clouds can span several orders of magnitude from 10$^{-6}$ to 10$^{-2}$ M$_{\odot}$ yr$^{-1}$. 
The well studied Orion A cloud is found to have a star formation rate of 7.15$\times$10$^{-4}$ M$_{\odot}$ yr$^{-1}$ \citep{Lombardi2010}. Recent studies of the massive star forming region G29.96−0.02 indicate a current star formation rate 0.001-0.008 M$_{\odot}$ yr$^{-1}$ \citep{Beltran2013}
which is also consistent with Galactic HII regions \citep{Chomiuk2011}. The mass of G29.96−0.02, a star-forming region in the Milky Way, is given as $\simeq$8$\times$10$^{4}$ M$_{\odot}$ \citep{Beltran2013} which will be be a useful comparison to our models. Our model will also be compared to the more massive 
($\approx$ 3$\times$10$^{5}$ M$_{\odot}$) G305 star forming cloud, located roughly 4 kpc away in the Scutum-Cruz arm of the Milky Way, with formation rates of 0.01-0.02 M$_{\odot}$ yr$^{-1}$ \citep{Faimali2012}.

\section{A basic model for cluster formation}

\begin{figure*}
 \begin{center}
  \includegraphics[width=.7\linewidth]{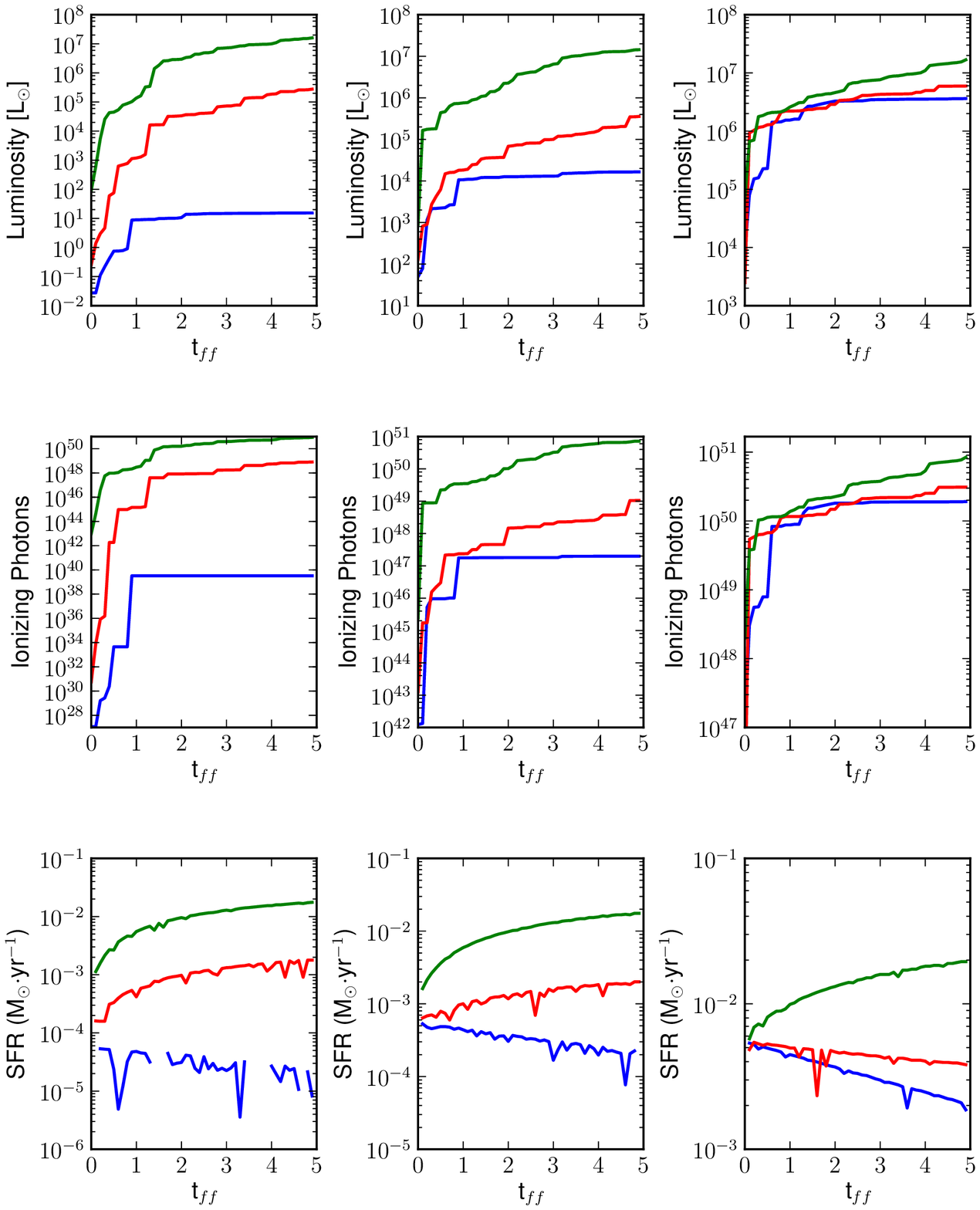}
 \end{center}
 \caption{A subset of our data showing the luminosity, number of ionizing photons, and the star formation rate. Blue, red, and green lines represent accretion rates of 0, 2.8$\times$10$^{-3}$ and 2.8$\times$10$^{-2}$ M$_{\odot}$ yr$^{-1}$, respectively. The 
 columns, from left to right, represent initial clump masses of 100, 1000, and 10000 M$_{\odot}$.}
\end{figure*}

\indent One of the most important aspects of radiative feedback of a young forming cluster on its surrounding host GMC is the shutting off of the accretion flow into the cluster forming region. To examine the radiative feedback effects of clusters on their surroundings, 
a cluster must be assigned the correct, combined radiative output of all its member stars as star formation proceeds. Cluster formation begins in a clump that has reached a critical density. Therefore, one input for a theory or subgrid model 
is how massive the original gas reservoir was at the moment that stars begin to form. Star formation proceeds as gas accretes onto this original dense region. Secondly, as its mass increases, the gravitational attraction of more material from the surrounding cloud will increase.   
The third step, ultimately, is the feedback from the cluster which helps to shut off the accretion flow onto the cluster forming clump. 

Therefore, the modeling needs to address two questions. How should the original gas reservoir be divided into stars? Second, how should the ongoing accreted gas be divided?  
The most straightforward way to address the former question is to divide the mass into stars at some prescribed efficiency according to an IMF. This ignores the effect of prestellar evolution but has the advantage of producing a cluster with the observed distribution
of stellar masses. While including prestellar evolution into the model may be more physically realistic, we argue that it is acceptable to place the stars directly on the main 
sequence in simulations that will run much longer than the prestellar evolution phase. Since higher mass stars evolve onto the main sequence rapidly, and the high mass stars
are the largest contributors in terms of ionizing feedback, ignoring prestellar evolution is a justified approximation. Further support comes from \cite{Klassen2012} who compared
the effects of radiative feedback from stars with prestellar evolution and those without it. The authors found that there was not a significant difference between the two cases. 

Since the cluster will be actively accreting gas until feedback effects stop the inflow, any added mass has to be dealt with accordingly. This gas can either be added to existing stars
or used to form new stars. While there are theoretical arguments for the accretion rate onto individual stars varying as M$^2$ \citep{Bondi,Krum2006,Throop} in the absence of turbulence, the fraction of gas accreted by existing stars
and the fraction of gas used to form new stars is unknown. To avoid adding uncertain parameters into our model, we therefore assume that a fraction of the available gas is only used to form new stars.

\begin{figure*}
 \begin{center}
  \includegraphics[width=.4\linewidth]{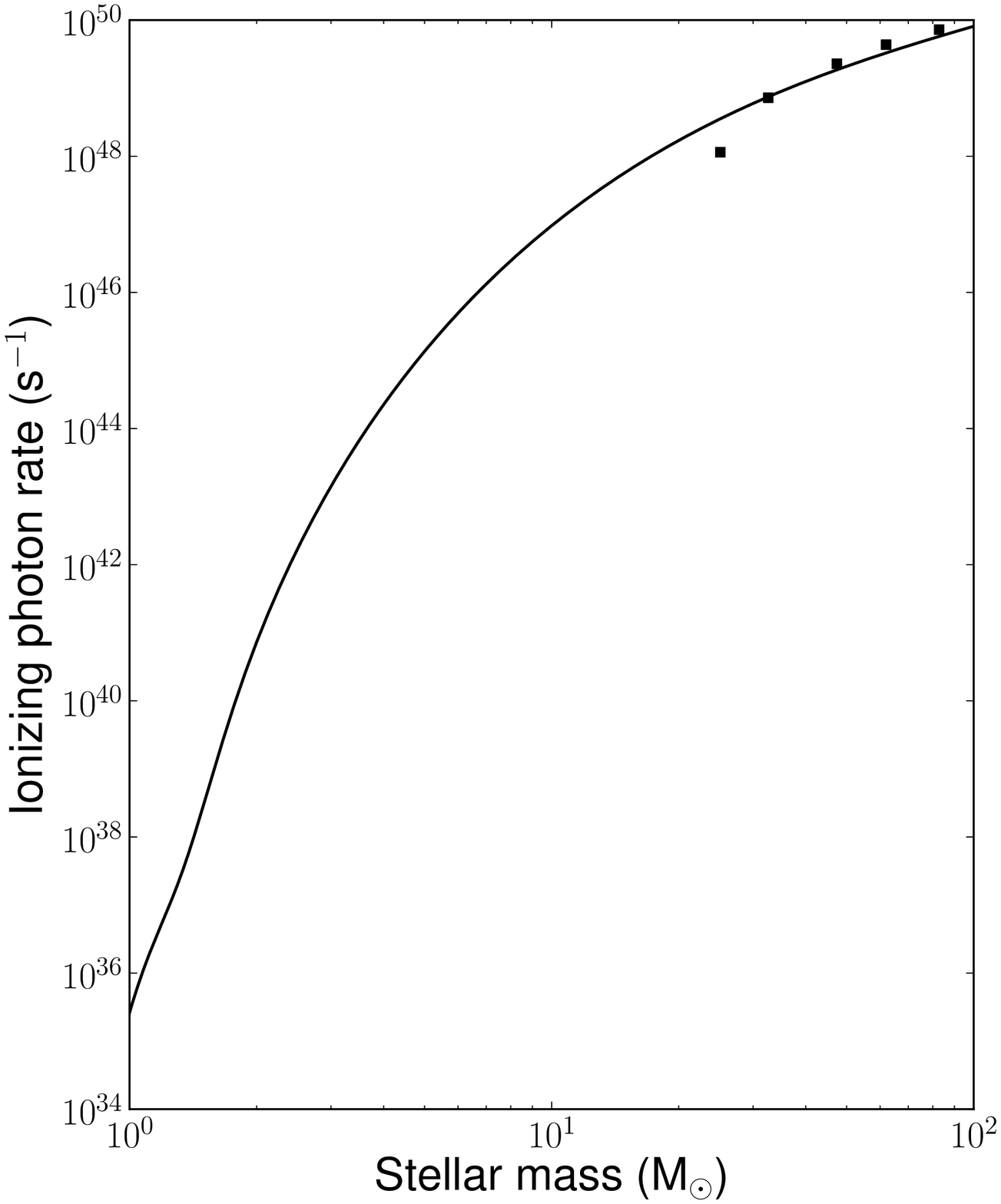}
 \end{center}
 \caption{The solid line represents the ionizing photons rate as a function of stellar mass found by directly integrating a blackbody function. The points represent results from \citet{Sternberg2003}.}
\end{figure*}

With this in mind, the concept of our model works as follows. A clump, assumed to be at the threshold density for cluster formation, forms and its mass is specified beforehand. From observational work, a number density of 10$^{4}$ cm$^{-3}$ is typical of a star forming clump \citep{Lada2003}. We therefore adopt this value as our density threshold for the formation of a cluster particle. 
The mass of the clump is initially divided into two categories; gas mass which will be used to form stars, and the leftover gas which will hereafter be designated as the 'reservoir'. 
The gas used for star formation is then distributed into main sequence stars via a randomly sampled IMF. Individual stellar masses are recorded so that the total mass of the clump in stars is known at any given time. 
If the random sampling of the IMF results in a star which has a mass greater than the total mass available for star formation then sampling is stopped and any remaining gas is added back to the reservoir. This process repeats by taking a fraction of the reservoir mass and converting it to stars. 
Over a sufficiently long time, all the mass in the cluster will be in the form of stars. Any mass accreted by the clump, with an accretion rate specified by the user, is added to the reservoir. It should be noted that accretion in this context, and throughtout the rest of the paper, refers to the replenishing of clump mass rather than
accretion onto individual protostars. 

Our star forming clumps are allowed to grow in mass through accretion but no mass loss is included. A physical motivation for our gas reservoir is required to justify this assumption. We posit that the reservoir gas in our subgrid model is in a dense enough state to remain bound
over long timescales even in the presence of stellar feedback. This has indeed been shown in simulations. \cite{Dale2005} showed that the inclusion of radiative feedback into cluster formation simulations resulted in collimated ionized outflows which are released perpendicular to the 
dense filaments from which they are forming. Even though these outflows accelerate a small fraction of the gas to high velocities, this is $not$ sufficient to unbind the bulk mass of the clump. In the case of external ionizing radiation, as discussed in \cite{Dale2007}, no dense star forming cores
are disrupted. There is also evidence for triggered star formation in these simulations suggesting that even if the gas is disrupted due to stellar feedback, it may still form stars in a separate region of the cloud. We therefore assume our gas reservoir is in a clumpy and filamentary
state, preventing its disruption via stellar feedback, meaning it is all available to form stars.  

To populate the cluster with stars, an IMF is randomly sampled with a Metropolis-Hastings algorithm. This algorithm is an example of a Markov Chain Monte Carlo method and works by generating a random walk and uses a specified probability distribution to either accept or reject the proposed move. More specifically, the acceptance ratio, $\alpha$, 
is calculated which is the ratio of probabilities between the proposed move and the previously accepted move. The acceptance ratio is then interpreted as the probability that the move is accepted (if $\alpha \geq$ 1 it is accepted automatically). We use the Chabrier IMF \citep{Chabrier2005} as our input probability distribution which is expressed as,

\begin{equation}
\xi(log \ m) = 
\begin{cases} 
0.093\times exp\{\frac{-(log \ m - log \ 0.2)^2}{2\times(0.55)^2}\}, & \mbox{\ m $\leq$ 1 \ M$_{\odot}$ } \\
0.041m^{-1.35 \pm 0.3}, & \mbox{\ m $>$ 1 \ M$_{\odot}$}.
\end{cases}
\end{equation}
\newline

\noindent Randomly sampling the IMF introduces stochastic effects into our model. The stars are allowed to have masses between 0.01 and 100 M$_{\odot}$. The lower limit is below the brown dwarf limit but we are concerned 
with the effects of the radiation field and brown dwarfs will not contribute significantly to the overall luminosity. The luminosity of a star is based on its mass via the function found in \cite{Tout1996}.

The star formation efficiency of 20$\%$ per freefall time is used in order to control the rate of star formation \citep{Lada2003}. The freefall time is given by,
\begin{equation}
t_{\rm{ff}} = \sqrt{\frac{3\pi}{32 G\rho}} = 0.36 (\frac{n}{10^4 cm^{-3}})^{-1/2} Myr
\end{equation}

\noindent which results in a value of 0.36 Myr assuming the density threshold of 10$^4$ cm$^{-3}$ discussed earlier. It should be noted that our star formation efficiency only reflects what is happening inside the cluster forming clump. 
The majority of the gas in a molecular cloud is at much lower density (on average about 100 particles per cm$^{-3}$) and may not end up in clusters meaning that the global star formation efficiency can be much less than 20$\%$. Since the IMF is not necessarily being sampled only once per freefall (see below), the fraction of 
total gas that is converted to stars every timestep is given by $f = M_{\rm{res}} \epsilon n/M_{\rm{tot}}$ where $M_{\rm{res}}$ is the reservoir of gas inside the cluster which has not been converted to stars, $\epsilon$ is the fraction of reservoir gas converted to stars per freefall time, $n$ is the number of times sampled in a freefall time, and M$_{\rm{tot}}$ is the total mass of the clump and cluster. 
The mass of all stars formed is tracked so that at any given time we know how much mass is tied up in stars and how much gas is available for future star formation. 

Recording the masses of all stars formed allows the ensemble properties of the accreting clusters to be determined. The total cluster luminosity is the sum of the inidividual stellar luminosities which are determined through the analytic formulas provided by \cite{Tout1996}. These formulas provide the temperature and radius
of main sequence stars from their masses which is then used to calculate the stellar luminosity. We also calculate the the ionizing photon rate, in s$^{-1}$, from the cluster which is again the sum of the individual stellar rates. The ionizing photon rate is found by directly integrating a blackbody distribution.  

\begin{figure*}
 \begin{center}
  \includegraphics[width=.8\linewidth]{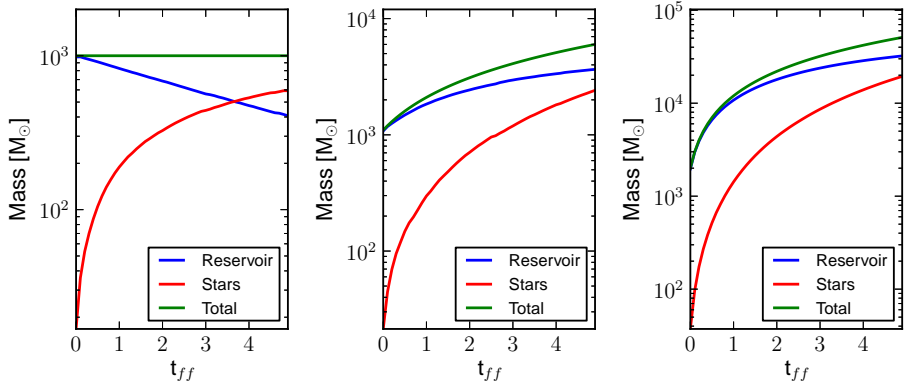}
 \end{center} 
 \vspace{2pt}
 \caption{The total gas mass, the mass in stars, and the reservoir mass as a function of time for an initial clump mass of 1000 M$_{\odot}$. The panels from left to right represent models with an accretion rate of 0, 2.8$\times$10$^{-3}$ M$_{\odot}$yr$^{-1}$,
 and 2.8$\times$10$^{-2}$ M$_{\odot}$yr$^{-1}$.}
\end{figure*}

How often the IMF is sampled for building new stars in a cluster forming clump can have a significant impact on the cluster's properties and evolution. If sampling is done infrequently (ie. a small $n$) then a larger amount of gas available for star formation will be converted
to stars, but the luminosity will have large and discontinuous jumps. On the other hand, if the IMF is sampled too frequently then the amount of gas that is being converted to stars each timestep
will be small meaning that high mass stars cannot be formed. This would lead to a bias towards low mass stars. It is therefore necessary to find the smallest value of
$n$ which still reproduces the IMF. 

To examine the effect of changing the IMF sampling rate, multiple models were run over one freefall time with varying initial clump masses
and accretion rates. We stress that the accretion rates are constant over time and were chosen to be representative of the actual accretion rates onto clusters. In a real cluster forming environment, the resulting
radiation field would eventually act to reduce the accretion rate by feedback. This step is handled in our full simulations of feedback onto the surrounding GMC gas. The results are
shown in Figure 1. The plots show the number of low mass stars ($<$ 1 M$_{\odot}$) and high mass stars ($>$ 1 M$_{\odot}$) as a function of the sampling rate. The number
of stars is shown relative to their expected numbers. The expected number of stars is found by directly integrating the Chabrier IMF for a cluster which contains the same mass in stars. A subset of points were ran 100 times each and the resulting average and two standard deviations are shown. It can be seen that the number of stars deviates
from the expected numbers at high sampling frequencies in all cases. The divergence point, however, does not occur at the same place and increases in sampling frequency
with increasing mass. This is easily understood in terms of the available mass used to form stars at the time of sampling. As the sampling rate increases, the amount of gas
being converted to stars decreases.

It can be seen from the figure that there is a large amount of scatter in the 100 M$_{\odot}$ case. This is a stochastic effect of the smaller
number of total stars formed in this case. Therefore, the formation of a single high mass star has a larger impact in comparison to the higher clump mass cases. 

Based on Figure 1, we have chosen to sample the IMF 10 times per freefall time which corresponds to approximately 36 kyrs. This sampling rate still reproduces the number of 
high and low mass stars for all mass ranges within error. As shown in the next section, this sampling rate also reproduces the correct number of O stars which are the most important
stellar population with respect to ionizing feedback. It should be noted that the chosen sampling time is less than the time it takes for a massive star to form and reach
the main sequence. If the sampling frequency was greater than this time then the ionizing luminosity of the cluster would be underestimated between samplings. 

\begin{figure*}
 \begin{center}
  \includegraphics[width=\linewidth]{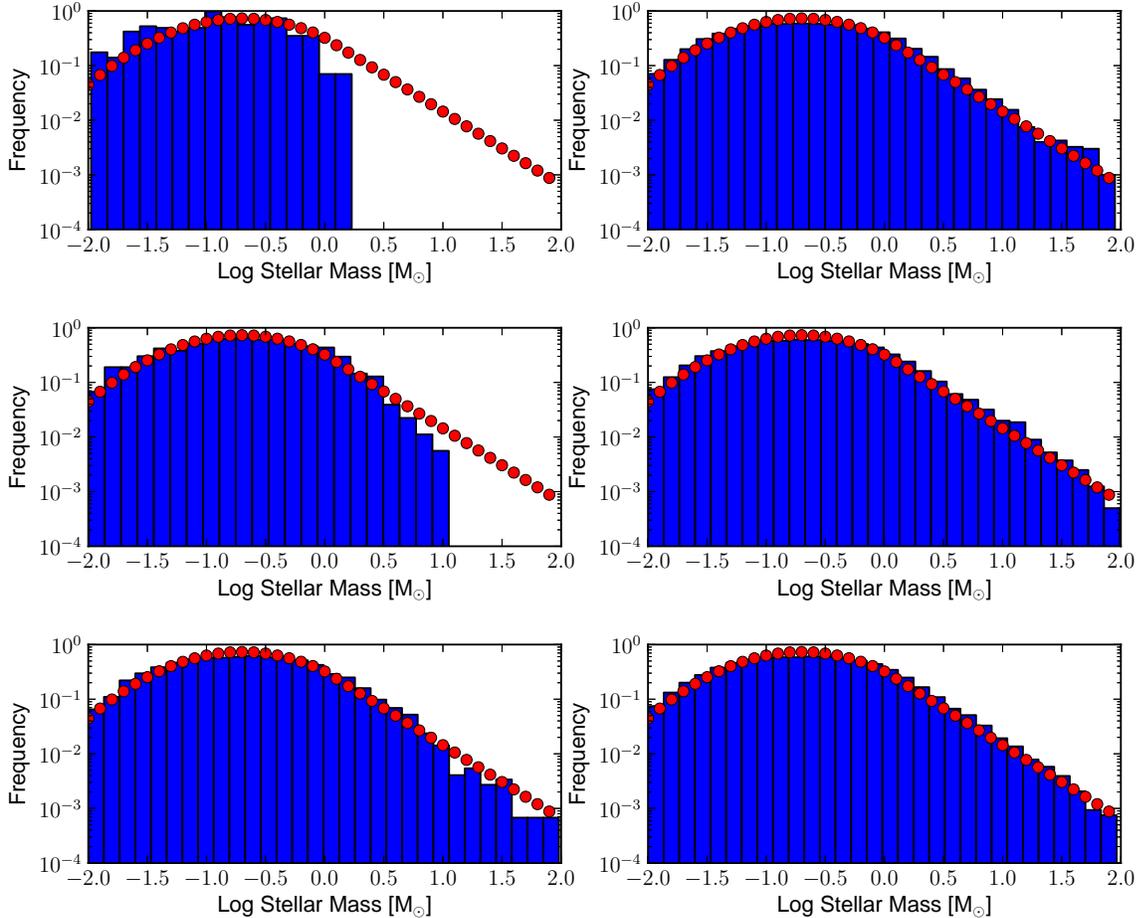}
 \end{center}
 \caption{The resulting mass function for models with no accretion (left) and an accretion rate of 2.8$\times$10$^{-2}$ M$_{\odot}$ yr$^{-1}$ (right) for initial clump masses of 100 (top), 1000 (middle), and 10000 (bottom) M$_{\odot}$. The red circles represent the normalized
 Chabrier IMF.}
\end{figure*}

In the following section, we present the result of a suite of models that were run with different initial cluster masses and different accretion rates.

\section{Results}

To examine the longterm behaviour of our model, we ran multiple simulations for various initial clump masses and accretion rates. All models were run for 5 freefall times or $\simeq$1.8 Myr. The initial clump masses ranged between 0 and 10$^5$ M$_{\odot}$. At 0 M$_{\odot}$, the
evolution is tied only to the mass that it accretes. The high initial clump mass of 10$^5$ M$_{\odot}$ was chosen to be representative of a protoglobular cluster. 

The accretion rate into the clump was chosen to be constant and had values ranging from
0 M$_{\odot}$t$_{\rm{ff}}^{-1}$ to 10$^{5}$ M$_{\odot}$ t$_{\rm{ff}}^{-1}$ (or $\approx$ 2.8$\times$10$^{-1}$ M$_{\odot}$ yr$^{-1}$). These rates were chosen to be representative of realistic cluster accretion rates. Observational studies have shown that
high mass protostars have accretion rates up to 10$^{-4}$-10$^{-3}$ M$_{\odot}$ yr$^{-1}$ \citep{Fuller2005,Beltran2006}. Therefore, our highest accretion rate roughly corresponds to a cluster which is forming multiple large mass stars. It should be noted that 
a cluster will not have a constant accretion rate especially over 5 freefall times. However, a constant accretion rate is the easiest to implement and can still provide information on how physical properties of our clusters vary depending on accretion rate. When this
model is used in hydrodynamical simulations, the accretion will be determined by the environment surrounding the cluster and not put in manually as is done here.

To ensure that our model is behaving as expected, we chose a subset of our data and plotted (see Figure 2) the luminosity, number of ionizing photons, and the star formation rate (SFR). These quantities were chosen as useful comparisons to observational data. The initial clump masses in Figure 2 are 100, 1000, and 10000 M$_{\odot}$ from left to right. The accretion rates
shown in blue, red, and green are 0, 2.8$\times$10$^{-3}$ M$_{\odot}$ yr$^{-1}$, and 2.8$\times$10$^{-2}$ M$_{\odot}$ yr$^{-1}$, respectively. 

The initial masses and accretion rates were chosen to represent a wide range of clusters. The final clump masses cover a range from 100 M$_{\odot}$ to 60000 M$_{\odot}$, the range of luminosities is 10 L$_{\odot}$ to 10$^{7}$ L$_{\odot}$, and ionizing fluxes ranging from 10$^{40}$ 
to 10$^{51}$ s$^{-1}$. The SFRs at low cluster masses show significant variability which can be attributed to the stochastic sampling of the IMF. The discontinuous lines in the SFR plots indicate that no star formation has occured during those timesteps.

There are two general trends to note in Figure 2. The first is that a higher accretion rate results in higher values for all three quantities plotted, as expected. Second is that as the initial clump mass increases the final luminosity and number of ionizing photons begin to converge. This is due to the large initial gas reservoir available which outweighs the effect of the smaller accreted mass. 

Since our model will be used to represent the radiative feedback of clusters, it is important to verify that the expected number of ionizing photons are being produced. The number of ionizing photons (E$_{\rm{phot}} >$ 13.6 eV) for clusters has been calculated in previous population synthesis models which include the 
effects of stellar evolution and metallicity \citep{Smith2002,Sternberg2003}. Our model reproduces their cited values to within an order of magnitude for similarly sized clusters. For example, an impulsive burst of star formation with a total mass of 10$^5$ M$_{\odot}$ and an upper IMF limit of 120 M$_{\odot}$ results in an ionizing photon rate of $\sim$8$\times$10$^{51}$ s$^{-1}$ in Sternberg et al. (2003). 
As a reference, we have shown in Figure 3 the ionizing photon output as a function of stellar mass for our model, found through integrating a blackbody, and overplotted results from \cite{Sternberg2003}. The large jumps in the number of ionizing photons can be traced back to the emergence of individual massive stars. For example, the steep rise in the model with an initial mass of 1000 M$_{\odot}$ and no accretion (blue line in centre panel of Figure 2) is due to the formation of a 11.2 M$_{\odot}$ star. It should be noted that our model does not include stellar deaths so the luminosity and number of ionizing photons can only increase with time.

The star formation rates shown in Figure 2 span a few orders of magnitude from 10$^{-5}$ to 10$^{-2}$ M$_{\odot}$ yr$^{-1}$. Recent studies of the massive star forming region G29.96−0.02 ($\approx$ 8$\times$10$^{4}$ M$_{\odot}$) indicate a current star formation rate 0.001-0.008 M$_{\odot}$ yr$^{-1}$ \citep{Beltran2013}. 
This roughly corresponds to our model with an initial mass of 10000 M$_{\odot}$ and an accretion rate of 10000 M$_{\odot}$ yr$^{-1}$ (green line in bottom right panel) which has a final mass of 60000 M$_{\odot}$ after 5 freefall times. The star formation rate for this case ranges from 0.005-0.02 M$_{\odot}$ yr$^{-1}$ and therefore agrees with the observation initially. 
Our model also agrees with the more massive ($\approx$ 3$\times$10$^{5}$ M$_{\odot}$) G305 star forming cloud with formation rates of 0.01-0.02 M$_{\odot}$ yr$^{-1}$ \citep{Faimali2012}. The RCrA star forming cloud is an example of a smaller region which has a mass of roughly 1100 M$_{\odot}$ and has a SFR of 2.5$\times$10$^{-5}$ M$_{\odot}$ yr$^{-1}$ \citep{Lada2010}. Our model with an initial mass of 1000 M$_{\odot}$ and no accretion (blue curve in middle, bottom panel) shows a SFR which is higher by an order of magnitude.
It should be noted that there is a wide range in SFRs between clouds with similar mass and since our model does not include accretion onto forming protostars, we provide an upper limit to the SFR. 

The evolution of the star formation rate with time shows two recurring trends: those with an increasing SFR, and those with a decreasing SFR. The difference between the two trends is due to the amount of gas in the reservoir. The cases showing a decreasing SFR have total accreted mass which is less than the initial
clump mass and is therefore reservoir dominated. This leads to a decreasing reservoir mass. Conversely, all cases which have an increasing SFR are accretion dominated. This leads to a build up of the reservoir mass and the corresponding increase in the SFR. As an example, consider
the middle panel in Figure 2 showing the SFR. This panel represents an initial clump mass of 1000 M$_{\odot}$. The case with zero accretion shows a decreasing SFR while the cases with $\dot{M}$ = 1000 and 10000 M$_{\odot}$ t$_{\rm{ff}}^{-1}$ have an increasing SFR. 

\begin{figure*}
 \begin{center}
  \includegraphics[width=\linewidth]{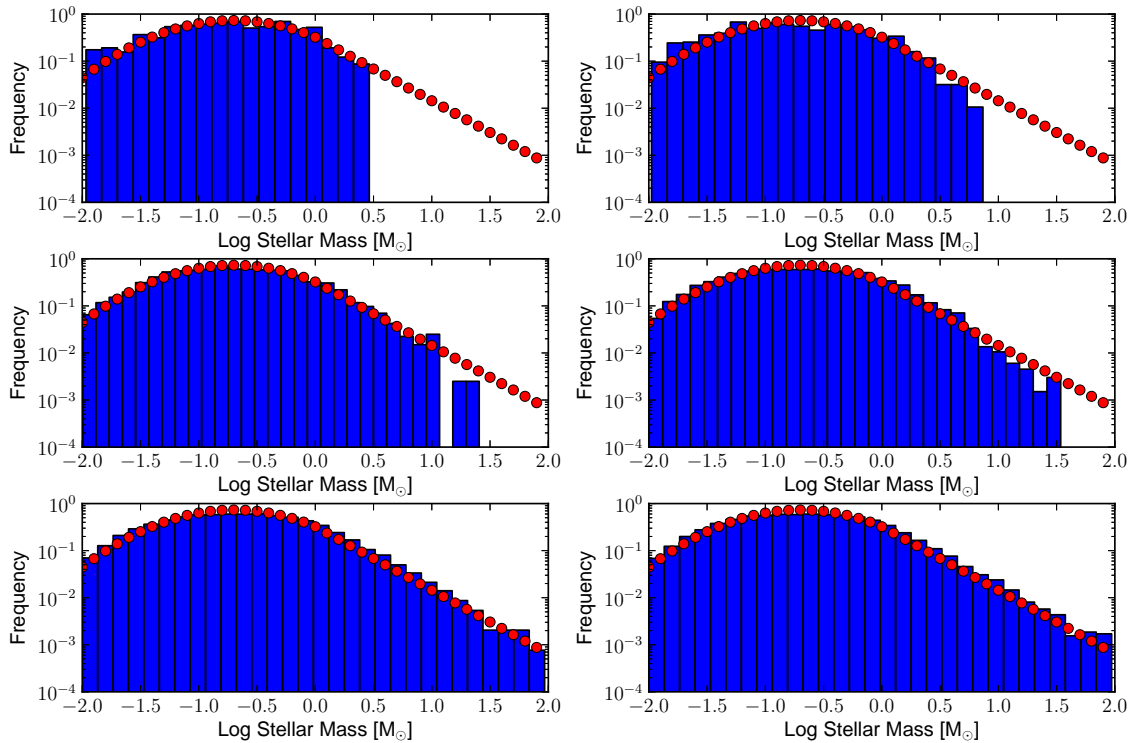}
 \end{center}
 \caption{The resulting mass function for models with the same final clump mass but have either accreted all the mass (left) or had no accretion and the mass was present initially (right). From top to bottom, the final clump masses are 500, 5000, and 50000 M$_{\odot}$. The red circles represent the normalized
 Chabrier IMF.}
\end{figure*}

Figure 4 shows how the total mass of the star forming region, the mass in stars, and the reservoir mass evolve with time. It can be seen that the SFR has qualitatively similar behaviour to the reservoir mass which confirms that the interplay between accretion and initial clump mass plays an important role in determining the cluster's properties.

\begin{figure*}
 \begin{center}
  \includegraphics[width=.7\linewidth]{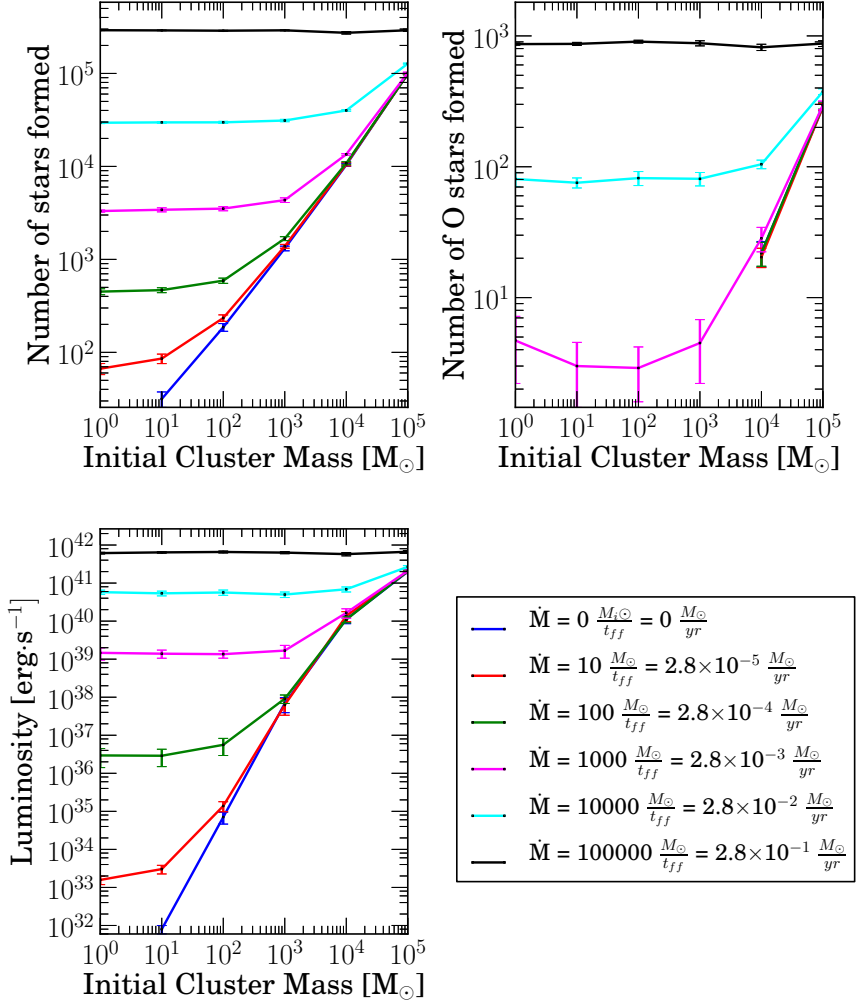}
 \end{center}
 \caption{The total number of stars, the number of O stars, and the final luminosity of runs with varying accretion rates and initial clump masses. The points represent the average of all runs and the error bars
 represent one standard deviation.}
\end{figure*}

The mass functions plotted in Figure 5 show that the powerlaw tail extends to higher masses when starting with a higher mass clump and no accretion. While the probability of forming a massive star is the same between cases, the extra condition that there must be enough available mass to form massive stars is responsible for the broadening of the mass function. The cases with an accretion
rate of 2.8$\times$10$^{-2}$ M$_{\odot}$ yr$^{-1}$ do not show this behaviour. This is because all models have accreted the same amount of gas and therefore show a similar mass function. Since this accretion rate is typical of actual clusters, this suggests that accretion into a clump may be partly responsible for the universality of the IMF since it provides enough mass to have a fully sampled IMF.

It is important to note that even though we are using the Chabrier IMF as a probability density function in this work, the resulting mass function will not necessarily be identical to the IMF. The imposed condition that there must be enough mass available during each sampling to form the randomly selected stars means that \emph{only the higher mass clumps can form massive stars}. In the case of lower mass clumps, massive 
stars cannot form so the high mass tail of the mass function is truncated at a value less than the maximum stellar mass of 100 M$_{\odot}$. The stars that do form, however, will follow the Chabrier IMF but the distribution will not entirely cover the allowed range of stellar masses. 

While Figure 5 shows that a large clump accretion rate results in a similar IMF independent of initial clump mass, it is difficult to draw further conclusions since all clumps shown have a different final mass. Therefore, we have plotted the resulting IMF of models which have the same final clump mass but either got their mass solely through accretion (left) or solely as the initial mass of the clump (right) in Figure 5. These represent the extremes of the accretion dominated and the reservoir dominated regimes.

The main difference between the two regimes is the extent of the high mass end of the IMF. The cases where all the mass was present in the initial clump forms more high mass stars. Conversely, cases where the majority of the clump mass is accreted have fewer high mass stars. We are only showing a representative case in Figure 6 but this trend holds in general. This result highlights the importance of how a clump or cluster gets its mass since we have shown that
the resulting IMF can differ depending on whether accretion is present. The difference can be attributed to the total mass in stars present at the end of the simulation. Reservoir dominated cases are found to have significantly higher masses in stars than accretion dominated cases. As an example, take the case with a total mass of 5$\times$10$^{4}$ M$_{\odot}$. The case with accretion has $\sim$19,000 M$_{\odot}$ in stars compared to $\sim$32000 M$_{\odot}$ for the case without accretion even though both clumps have the same total mass. This also translates to a lower SFR and a higher star formation timescale in 
accretion dominated cases. Radiative feedback may therefore play less of a role early on in the star formation process for low mass clumps which are actively accreting gas.

It is important to note that the differences between the \emph{reservoir dominated} and \emph{accretion dominated} regimes decrease with increasing clump mass. This is because there is enough gas present to have a fully sampled IMF regardless of how the mass was obtained (shown in the bottom panel of Figure 6).
\begin{figure}[t]
 \begin{center}
  \includegraphics[width = 0.85\linewidth]{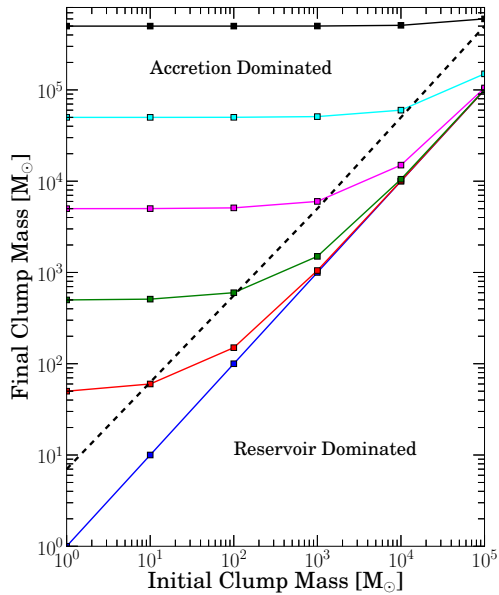}
 \end{center}
 \caption{A reference figure showing the final clump masses for the models run in Figure 7 with the same colour coding. The dotted black line separates the accretion and reservoir dominated regimes.}
\end{figure}
The analysis presented above is based on single model runs. Since there is a stochastic component related to the sampling of the IMF, it is useful to quantify how results vary between runs with identical input parameters.

In Figure 7, we plot the total number of stars formed, the number of O stars formed, the resulting luminosity, the total mass in stars, and the SFR for models with varying initial clump masses and accretion rates. The code was allowed to run for 5 freefall times as above. Models with accretion rates between 0 and 2.8$\times$10$^{-2}$ M$_{\odot}$yr$^{-1}$
were run 100 times each and the highest accretion rate of 2.8$\times$10$^{-1}$ M$_{\odot}$yr$^{-1}$ was run 10 times. The average values and corresponding standard deviations are plotted in Figure 7. It can be seen that our model covers a wide range of cluster types, ranging from
small clusters with less than 100 stars to the globular cluster regime. There is variation between runs with identical input parameters but the results are typically consistent to within a factor of 3. It can be seen from these plots that all three quantities
increase with accretion rate as expected. There is also a flat region that occurs at low initial clump masses for all quantities. The size of the flat region, however, increases with increasing accretion rate. 

Again, this can be understood in terms of the relative importance of initial clump mass and the accretion rate. The accretion dominated cases show little variation in the quantities. This is because the initial mass is a small fraction of the final mass 
so its effect is `washed out'. Only when the accreted mass becomes significantly smaller than the initial clump mass do the quantities begin to vary. Take, for example, the case with an accretion rate of 2.8$\times$10$^{-3}$ M$_{\odot}$ yr$^{-1}$.
The total accreted mass after 5 freefall times is 5000 M$_{\odot}$. All quantities remain constant for cases where the initial mass is $\leq$ 1000 M$_{\odot}$. There is an increase in all quantities when the initial clump mass
is 10$^4$ M$_{\odot}$, confirming that the interplay between initial clump mass and accreted mass is important in determining the cluster's properties.

To make the distinction between accretion and reservoir dominated regimes more clear, we have plotted the final clump mass versus the initial clump mass in Figure 8. This allows one to determine the properties given in Figure 7 for a region with a desired final mass. The dotted black line roughly shows
the transition between the accretion dominated and reservoir dominated regimes. Figure 8 is qualitatively similar to the plots shown in Figure 7 suggesting that these quantities scale directly with the clump mass. 

\begin{figure*}
 \begin{center}
  \includegraphics[width=0.8\linewidth]{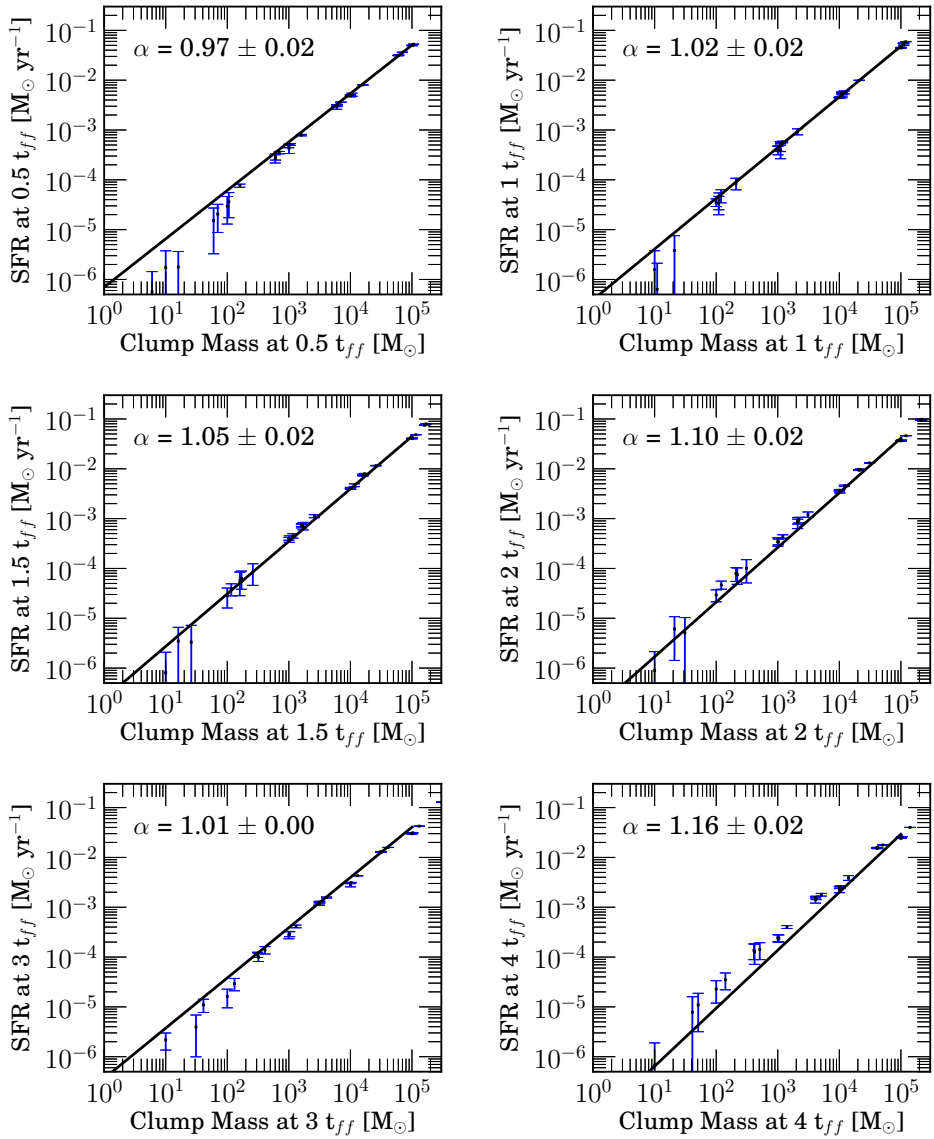}
 \end{center}
 \caption{The instananeous SFR as a function of total clump mass at 2 freefall times. Error bars represent the resulting standard deviations from 100 identical runs. The $\alpha$ parameters shown in the plots are the resulting indexes from powerlaw fits to the data. The error on
 $\alpha$, obtained through the fitting, is also shown.}
\end{figure*}

We have also examined how the SFR scales with clump mass. The SFR is chosen because it has been measured in a variety of star forming environments over scales ranging from individual clumps to entire GMCs. Rather than plotting the final clump mass, however, we have chosen to plot
the instantaneous SFR versus the clump mass at 0.5, 1, 1.5, 2, 3, and 4 freefall times to give snapshots of the evolution at several stages. The results for all accretion rates are shown in Figure 9. As in Figure 7, the error bars are the resulting
standard deviations from 100 runs with identical input parameters. The distribution was fit with a powerlaw given by,
\begin{equation}
SFR = const\cdot M^{\alpha}
\end{equation}

\noindent and the resulting index values shown in Figure 9 fall in the range

\begin{equation}
\alpha = 0.97-1.16.
\end{equation}

There are several studies which show that the observed SFR is linearly related to the dense gas mass. Within our own galaxy, \cite{Wu2005} examined the far infrared (FIR) luminosity versus the HCN luminosity in
nearby clouds. HCN requires dense gas (n $\geq$ 10$^4$ cm$^{-3}$) in order to be visible, and the FIR luminosity traces star formation. The authors found that there is a linear relation between the two quantities. This relation has also been found
in normal spiral and starburst galaxies by \cite{Gao2004}. These results are also supported by \cite{Alves2010} who found that the number of YSOs is linearly related to the gas mass above a given density threshold. The data suggests that our model can reproduce the 
behaviour of clumps and clusters with a wide range of physical characteristics.

Figure 9 also shows that the index $\alpha$ increases slowly with time, most likely due to the accelerating SFR in accretion dominated cases and the decreasing SFR in reservoir dominated cases.
The very low mass clumps are certainly reservoir dominated, and therefore have decreasing SFR with time, while the highest mass clumps are accretion dominated. The increasing SFR on the high mass end together with the
decreasing SFR on the low mass end result in a steepening of the slope.

While one might assume that if we convert a fixed amount of gas to stars (20$\%$ of the available reservoir gas per freefall time) then a linear relationship between the SFR and clump mass is a direct underlying result of our model. This is not generally true. There are reasons, however, why the linear dependence is unexpected and therefore significant.
Since we are converting 20$\%$ of the \emph{available reservoir gas} rather than 20$\%$ of the total cluster mass (reservoir gas plus stars), the SFR is sensistive to the accretion history of the cluster. This is most easily understood in terms of clusters that have identical masses
but are either in the extreme accretion dominated or reservoir dominated regime. In the accretion dominated regime, the amount of gas in the reservoir is increasing and, as shown earlier, so is the SFR. The opposite is true for the reservoir dominated regime. Therefore, two clusters with identical masses can have different amounts of reservoir gas which translates to different SFRs.
This could lead to a non-linearity in the SFR vs clump mass plots. The condition that, in a given timestep, there must be enough available gas to form the randomly selected star can also lead to non-linearities escpecially in the case of low mass clumps. It can be seen that there is a slight depature from the linear behaviour of the plots seen in Figure 9 below $\sim$ 100 M$_{\odot}$. These points are consistently 
below the line, likely due to the lack of massive star formation. Since these low mass clumps are not converting a significant amount of gas to stars each timestep, a large fraction of randomly selected stars cannot form. This provides a useful prediction for the SFR in small mass clumps; namely, there should be a departure from a linear dependence at low clump masses.

\begin{figure}
 \begin{center}
  \includegraphics[width=0.9\linewidth]{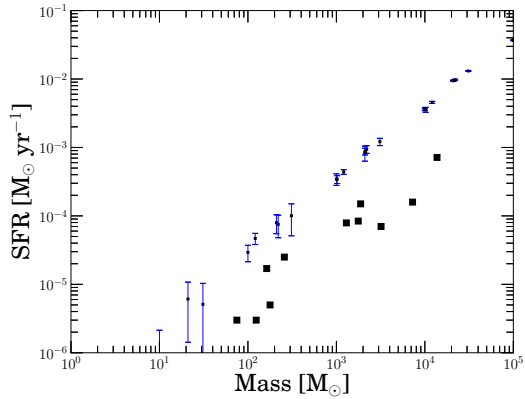}
 \end{center}
 \caption{The SFR versus clump mass for this work, shown in blue, and observations presented in \citet{Lombardi2010}, shown in black. The observed SFRs are systematically lower compared to our model which we attribute to feedback which alters the clump accretion rate and decreases
 the SFR.}
\end{figure}

As a further comparison to observed star-forming clumps, in Figure 10 we have plotted the SFR versus clump mass at 2 freefall times and overplotted the observed SFRs inferred by \cite{Lombardi2010} for nearby star-forming complexes. We have only shown our data at 2 t$_{\rm{ff}}$ because the same general
trend holds for the SFR at any time. We find that our data is consistently higher than the observed SFRs for all points and do not agree within error. Our data does agree with the measured values to within an order magnitude. We have attributed the difference in SFRs to the lack of 
feedback included in our subgrid model. We have shown previously that the accretion history of a clump can have a significant impact its SFR and feedback, especially radiative feedback, can greatly alter the accretion \citep{Dale2007,Peters2010,Bate2012,Dale2012,Klassen2012,Kim2012}. On smaller scales, the inclusion of radiative feedback suppresses fragmentation 
by increasing the temperature locally. On larger scales, ionizing radiation can clear gas from the clump by driving large scale outflows. These mechanisms tend to decrease the accretion rate and the SFR. When our model is included in full hydrodynamical simulations which include
radiative feedback, the accretion will be determined self-consistently rather than specified beforehand. The full simulations should therefore show better agreement with the measurements by \cite{Lombardi2010}.

The characteristic timescale $\tau_{\rm{SFR}}$ for star formation is by definition
$\tau_{\rm{SFR}} = M / (SFR) \sim M^{1-\alpha}$.  The actual model numbers from the 
different panels in Fig.~8 fall in the range $\simeq 3$ Myr within factors
of two.  Since $3 t_{\rm{ff}} \sim 1$ Myr in our simulations, our results should provide 
in principle provide a useful prediction for the
expected \emph{age range} within a young, relatively massive star cluster: they should
take only a few Myr to build.  Direct observational 
comparisons with such objects nearby are made more difficult by the inevitable presence of dust
and differential extinction within actively star-forming clusters, but an age range at the
3-Myr level seems comfortably realistic.  One example is the massive ($> 2 \times 10^4 M_{\odot}$)
cluster R136 in 30 Doradus, for which a recent study \citep{andersen2009} finds a mean 
age of 3 Myr and an internal age spread of nearly the same amount \citep{Massey1998}.
A very similar set of conclusions has been suggested for the massive young cluster at the
centre of NGC 3603 \citep{melena2008,pang2013}, 
in which an age range of a few Myr may show up for the lower-mass stars particularly.

\begin{figure}
 \begin{center}
  \includegraphics[width=0.8\linewidth]{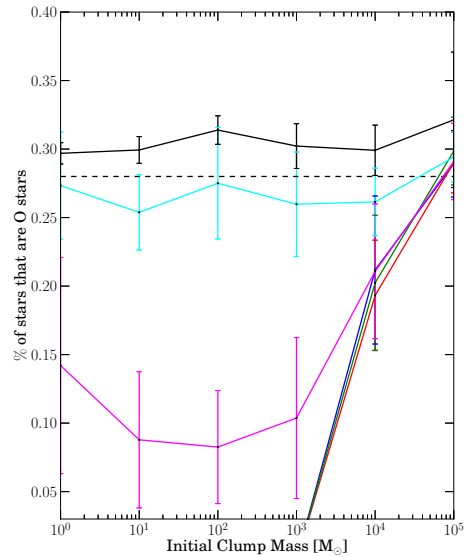}
 \end{center}
 \caption{The percentage of O stars formed from the models shown in Figure 7. As discussed in the text, the expected percentage of O star from directly integrating the IMF is 0.28$\%$. The points represent the average of all runs and the error bars
 represent one standard deviation. The colour coding is the same as in Figure 7.}
\end{figure}

One of the most important properties of a cluster in terms of its radiative feedback is the number of O stars that are formed. In this work, an O star is defined to be any star whose mass exceeds 16 M$_{\odot}$. Since a single O star can have a luminosity greater than 10$^5$ L$_{\odot}$ \citep{Carroll} and have most of its energy output in the form of
UV photons, under or over-producing O stars can significantly alter how the cluster interacts with its surroundings. It is therefore important to verify that our model is reproducing the expected number of O stars. Directly integrating the IMF results in approximately 0.28 $\%$ of the stars formed having masses greater than 16 M$_{\odot}$. In Figure 11, we have plotted the percentage of stars which are O stars. As in Figure 7, the points represent the average value and the error bars are one standard deviation.
It can be seen from this figure that the highest mass cases agree with the expected value. 

Figure 11 shows that the lowest mass model which produces O stars has a final clump mass of 5000 M$_{\odot}$ which gives a rough threshold for the emergence of massive stars. Interestingly, the present day mass of the ONC is roughly 4800 M$_{\odot}$ \citep{Hillenbrand1998} and also contains a small number of massive stars ($\sim$ 5 stars greater than 16 M$_{\odot}$ from \cite{Hillenbrand1997}) suggesting that our model is behaving as expected. 
The model with an accretion rate of 1000 M$_{\odot}$t$_{\rm{ff}}^{-1}$ seems to be underproducing O stars in the cases of low initial clump mass compared to what is expected directly from the IMF. Work by \cite{Williams1997} suggests that roughly 50$\%$ of clouds with a mass of 10$^5$ M$_{\odot}$ should form at least one O star which is
consistent with our results. This, together with the number of ionizing photons from Figure 2, suggests that our model is working as it should. 

There are a few caveats to our model. The expected percentage of O stars mentioned earlier assumes that the IMF is fully sampled which may not be the case if there is insufficient mass to form massive stars.
Also, the clumps in our model are still actively accreting gas after 5 freefall times so star formation is still ongoing and massive stars still have a probability of forming. This can be seen in Figure 4 from the large reservoir masses and in Figure 2 from the non zero SFRs at the end of the simulation.
The imposed condition of constant accretion onto the clump is also artificial. The actual clump accretion rate could be higher as suggested in \cite{Murray2012}. A higher accretion rate would lead to a building up of the reservoir mass leading to a higher chance of forming an O star. 

\section{Summary and Conclusions}

We have presented a model of cluster formation which is simple and general enough to be used in hydrodynamical simulations to represent the radiative output of an entire star cluster. The model starts with a clump of a specified mass and forms main sequence stars by sampling the Chabrier IMF. Every tenth of a freefall time, or 3.6$\times$10$^{4}$ years 
assuming a density threshold of 10$^4$ cm$^{-3}$ for cluster formation, a fraction of the remaining gas inside the cluster is converted to stars. We have shown that sampling every tenth of a freefall time reproduces the correct number of high and low mass stars. To test our model, we ran multiple simulations by varying the initial clump size and constant accretion rate. It should be noted that our model does not include the effects of radiative feedback as this will be handled in full hydrodynamical simulations. 
The point of our work is to provide a set of baseline models for how a clump converts into stars for a given accretion rate. The inclusion of radiative feedback would only change the cluster accretion rate. The inner workings of the clump/cluster, however, will remain the same. 
Therefore, the difference between the results in this paper and the properties of clusters in full hydrodynamical simulations may not be significant. With this in mind, we find the following:

\begin{itemize}
 \item The model we have presented is a straightforward way to determine the radiative output of clusters ranging in size from small subclusters of $\sim$10 M$_{\odot}$ to the globular cluster regime ($\sim$ 5$\times$10$^5$ M$_{\odot}$) with varying accretion rates.
 \item How a clump gets its mass has an impact on its properties. There are two different regimes we have identified; 'accretion' dominated, and 'reservoir' dominated. The accretion dominated
 regime is characterized by an increasing SFR and a less sampled IMF due to the smaller number of stars formed. The reservoir dominated regime has all mass present in an initial clump
 and is characterized by a descreasing SFR and a more fully sampled IMF. Differences between the two regimes disappear with a sufficiently large clump mass.
 \item We have shown that typical clump accretion rate can produce a fully sampled IMF regardless of the initial clump mass at the onset of star formation.  
 \item Our model reproduces the number of ionizing photons released from a cluster from previous, more detailed simulations. For a clump size on the order of 10$^4$ M$_{\odot}$ the resulting ionizing photon rate is between 10$^{50-51}$ s$^{-1}$. For high mass clusters, this is expected because the IMF is fully sampled.  
 \item We find that small ($\sim$ 100 M$_{\odot}$) clusters have SFRs of 10$^{-4}$-10$^{-5}$ M$_{\odot}$ yr$^{-1}$ while larger clusters ($>$ 10$^{4}$ M$_{\odot}$) can exceed 10$^{-2}$ M$_{\odot}$ yr$^{-1}$. Lower mass clusters also exhibit more variablility in their SFRs than higher mass clusters.
 \item The SFRs found in our model are systematically lower than those found in \cite{Lombardi2010}. This is because we deliberately ignore the radiative feedback on the medium surrounding the clump which would suppress the cluster accretion rate. As shown above, this will alter the star forming properties of the cluster.
 \item The SFR is nearly proportional to clump mass at all times $\leq$ 4 t$_{\rm{ff}}$. This agrees with multiple observations of star forming clump on both large and small scales \citep{Gao2004,Wu2005,Alves2010,Lombardi2010}.
 \item High mass clusters ($>$ 10$^{4}$ M$_{\odot}$) produce the correct number of O stars expected from the IMF. We have identified a final clump mass threshold for O star formation of $\sim$5000 M$_{\odot}$ which consistent with observations of 5 stars with masses greater than 16 M$_{\odot}$ in the ONC \citep{Hillenbrand1997} which has a present day mass of $\sim$4800 M$_{\odot}$.
\end{itemize}

\section*{Acknowledgments}

We thank the referee, Robi Banerjee, for insightful comments which helped improve this paper. R.E.P. ~and W.E.H ~are supported by Discovery Grants from the Natural Sciences and Engineering Research Council (NSERC) of Canada. This work was made possible by the facilities of the Shared Hierarchical Academic Research Computing Network (SHARCNET: www.sharcnet.ca) and Compute/Calcul Canada.

\bibliography{howardpaper.bib}
%\bsp

\label{lastpage}

\end{document}